\documentclass{article}

\usepackage{graphicx}
\usepackage{amsmath}
\usepackage{fullpage}
\usepackage{authblk}
\usepackage[hidelinks]{hyperref}

\newcommand{\dd}{\mathrm{d}}
\newcommand{\wlc}{\ensuremath{\omega_{LC}}}
\newcommand{\wm}{\ensuremath{\Omega_\mathrm{M}}}

\title{Dynamical backaction in nanoscale superfluid electromechanics}

\author{Marek Tal\'{i}\v{r} \thanks{marek.talir@matfyz.cuni.cz}}
\author{Filip Novotn\'{y}}
\author{Bal\'{a}zs Szalai}
\author{Nasiruddin Mondal}
\author{Emil Varga \thanks{emil.varga@matfyz.cuni.cz}}
\affil{Faculty of Mathematics and Physics, Charles University, Ke Karlovu 3, 121 16, Prague, Czech Republic}
\date{}

\begin{document}

\maketitle

\begin{abstract}
Nanofluidic acoustic resonators employing superfluid $^4$He can be used to study quantized vortices from collective behavior in two-dimensional superfluid turbulence down to few individual vortices created by rotation. In order to improve sensitivity to the level needed for probing individual quantized vortices, readout mechanisms employing optomechanical or optomechanics-inspired approaches seem to be promising in this regard. In this work, we develop an electromechanical system, which couples a 4$^\mathrm{th}$ sound acoustic resonance to a superconducting LC tank circuit in a sideband-resolved regime. Using this system, we demonstrate electromechanically induced transparency (EMIT), optical spring effect and optomechanical sideband damping and amplification. Furthermore, by rotating the cryostat, we demonstrate sensitivity to quantized vortices, which can become trapped and released in avalanche-like process inside the nanofluidic volume.
\end{abstract}


\section{Introduction}
\label{sec:intro}
Parametric coupling of a mechanical resonator to an electromagnetic resonance, optomechanics \cite{Aspelmeyer2014}, provides a versatile tool set for sensing and controlling mechanical motion, enabling, e.g., highly sensitive force sensing \cite{gavartin_hybrid_2012} or fundamental tests of quantum mechanics \cite{kotler_direct_2021}. Optomechanical systems based on superfluid helium are in particular interesting due to natural compatibility with cryogenic systems and high mechanical quality factors \cite{Spence2021} with possible applications for detection of continuous gravitational waves \cite{de_lorenzo_ultra-high_2017,Vadakkumbatt2021} or ultralight candidates of dark matter \cite{hirschel_superfluid_2024}.

Nanofluidic Helmholtz resonators, where acoustic motion of superfluid helium modulates a compliant capacitor, were demonstrated to be a viable mechanical element in microwave cavity electromechanics \cite{spence_three-tone_2023} and  feedback control of the mechanical motion \cite{varga_electromechanical_2021}. In this work, we continue this development and couple the Helmholtz resonator to a radio frequency LC tank circuit. The resulting electromechanical system is sideband-resolved and, using electromechanically-induced transparency and absorption (EMIT/A) \cite{Aspelmeyer2014}, we demonstrate dynamical backaction, i.e., optomechanical damping and anti-damping and the optical spring effect.

Of particular interest is that the superfluid $\mathrm{4^{th}}$ sound \cite{Tilley_book} present in the nanofluidic confinement above approximately 1 K, i.e., a coupled wave of pressure and entropy density with stationary viscous normal fluid, dissipatively couples to the motion of quantized vortices present in the system, which can be used for their detection \cite{novotny_detection_2024}. Quantized vortices, i.e., topological defects of the superfluid order parameter \cite{Tilley_book}, can be injected into the nanofluidic cavity either via excitation of turbulence \cite{varga_observation_2020,novotny_critical_2025,novotny_turbulent_2025,novotny_decay_2026} or by rotation \cite{novotny_detection_2024}.

Quantized vortices in such confined geometry are subject to strong pinning on the irregularities of the inner surface \cite{novotny_decay_2026}. Therefore, the distribution of the vortices in the system does not necessarily follow the equilibrium, given by instantaneous angular velocity of the system, but can become trapped in metastable states, among which it can switch in an avalanche-like reorganization. This process is believed to play an important role in the so-called "glitches" exhibited by several known pulsars \cite{melatos_avalanche_2008,wlazlowski_vortex_2016,warszawski_unpinning_2012,liu_vortex_2025}. In this work, we also demonstrate advantageous combination of sideband anti-damping and EMIA for fast readout of the dissipation of the acoustic mode, allowing us to resolve in time sudden changes in vortex-induced dissipation upon sudden disturbance of the system, in qualitative agreement with avalanche-like picture of vortex motion.

The paper is organized as follows: In Section~\ref{sec:setup} we introduce the nanofluidic devices and their parametric coupling to an LC tank circuit; in Section~\ref{sec:backaction} we demonstrate EMIT/A and dynamical backaction in this system and in Section~\ref{sec:vortices} we show the response of the dissipation of the mechanical acoustic mode to rotation and flow history of the device, after which the conclusions follow.

\section{Experimental Setup}
\label{sec:setup}
\begin{figure}
    \centering
    \includegraphics[width=\linewidth]{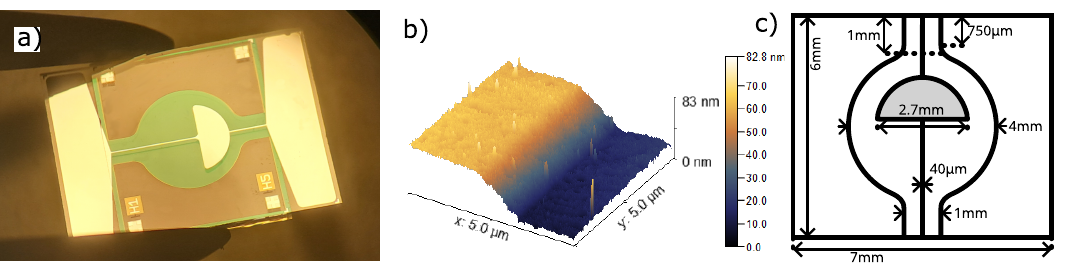}
    \caption{The nanofluidic Helmholtz resonator. a) Photograph of the resonator. Green colour is due to thin-film optical interference and outlines the nanofluidic cavity. b) Atomic force micrograph of the electrode edge inside the basin. c) Schematic view of the resonator dimensions, electrodes in gray. Electrode thickness is approximately 60 nm, cavity height quartz-to-quartz 900 nm.}
    \label{fig:fig1}
\end{figure}

The electromechanical system was realized using a nanofluidic Helmholtz resonator shown in Fig.~\ref{fig:fig1} wired in a parallel LC resonant circuit, shown in Fig.~\ref{fig:fig2}b, where the nanofluidic Helmholtz resonator acts as the variable capacitance coupled to the helium motion. The parallel LC circuit connects to a strip line terminated with SMA connectors, via a coupling capacitor ($C_\mathrm{ex}$ in Fig.~\ref{fig:fig2}b) on one end and is connected to a ground plane on the other.

The Helmholtz resonator shown in Fig. \ref{fig:fig1}, similar to one used in \cite{novotny_detection_2024}, is made of two amorphous SiO$_2$ chips in which a channel is etched, joined together by direct bonding to form a 900~nm high enclosed cavity. Electrodes are sputtered on the cavity walls, and patterned by lift-off to match an acoustic pressure maximum of an antisymmetric superfluid acoustic mode of interest, so that a resonant superfluid displacement changes the electrode capacitance. The electrodes are made of 60 nm thick layer of sputtered Nb (Fig.~\ref{fig:fig1}b shows AFM scan of the edge of the electrode), superconducting transition of which we measured to be well above 4~K. The capacitance of the devices is $C \approx 25$~pF.

The inductance in the LC circuit is a superconducting coil with approximately 200 windings of formvar-insulated 40~\textmu m single-filament NbTi wire wound on a 1~cm diameter PTFE core (16~mm long, around 10~mm covered with windings). The windings are tightly spaced, held in place by varnish and tightly wrapped in PTFE tape. The winding density is around 25~mm$^{-1}$ and the estimated total inductance is about 290~nH. The coupling capacitor C$_\mathrm{ex} = 10$~pF is a standard SMD capacitor with NP0 dielectric.

For the experiment, the LC circuit was submerged in superfluid helium in a bath cryostat, with base temperature of approximately 1.20~K. The entire cryostat and detection electronics were placed on a rotating platform \cite{novotny_detection_2024,dwivedi_dynamics_2024} capable of angular velocities up to 180 $^\circ$/s, and the angular velocity was measured by a commercial miniature MEMS gyroscope mounted on the platform.

\subsection{LC and acoustic resonance characterisation}
When submerged in superfluid helium, fourth sound acoustic resonant modes exist in the channel geometry of the Helmholtz resonator. The tight confinement of the channel permits only the superfluid component to move, leaving the normal fluid viscously clamped \cite{rojas_superfluid_2015,Souris2017}. The coupled system of the LC circuit and Helmholtz resonator can be well described by the theory of cavity optomechanics \cite{Aspelmeyer2014}, as the microfluidic resonance amplitude directly changes the Helmholtz resonator capacitance, which in turn shifts the LC resonance frequency. 

\begin{figure}
\includegraphics{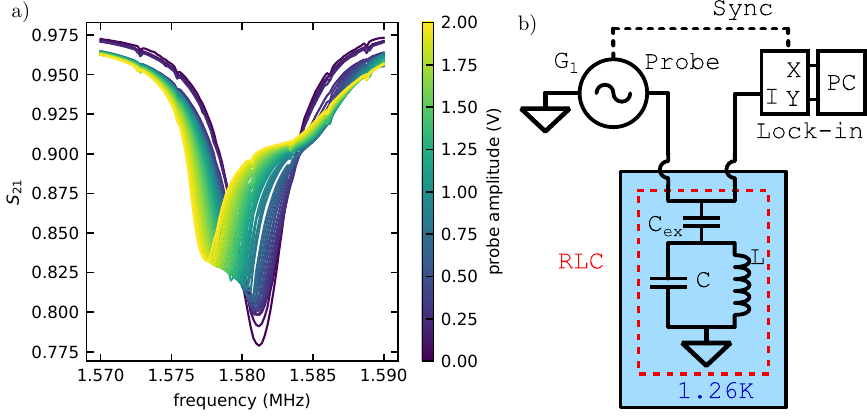}
\caption{Resonance of the LC tank circuit. a) Normalized transmission of LC resonance with increasing probe amplitude, showing nonlinearity and $Q$ degradation above 0.5~V. b) The circuit diagram for measuring the LC resonance.}\label{fig:fig2}
\end{figure}

The resonance of the LC circuit can be seen in Fig. \ref{fig:fig2}a for increasing probe amplitudes, measured at 1.26 K. The transmission is normalized to far-off-resonance response. The resonance frequency at the lowest amplitude is $\omega_{C} = 2\pi \times 1.5812$~MHz, with the total LC decay rate $\kappa = 2\pi \times 4.62$~kHz and external coupling of $\kappa_\mathrm{ex} = 2\pi \times 966$~Hz, making the quality factor $Q_{LC} = 340$. Despite the relatively low $Q_{LC}$, the total LC decay rate is sufficiently low so that the acoustic mode of interest (around 23 kHz, see below) is sideband resolved. Increasing probe amplitude lowers the cavity resonance frequency and $Q$ and makes the peak increasingly non-linear. 

\begin{figure}
    \includegraphics[width=\linewidth]{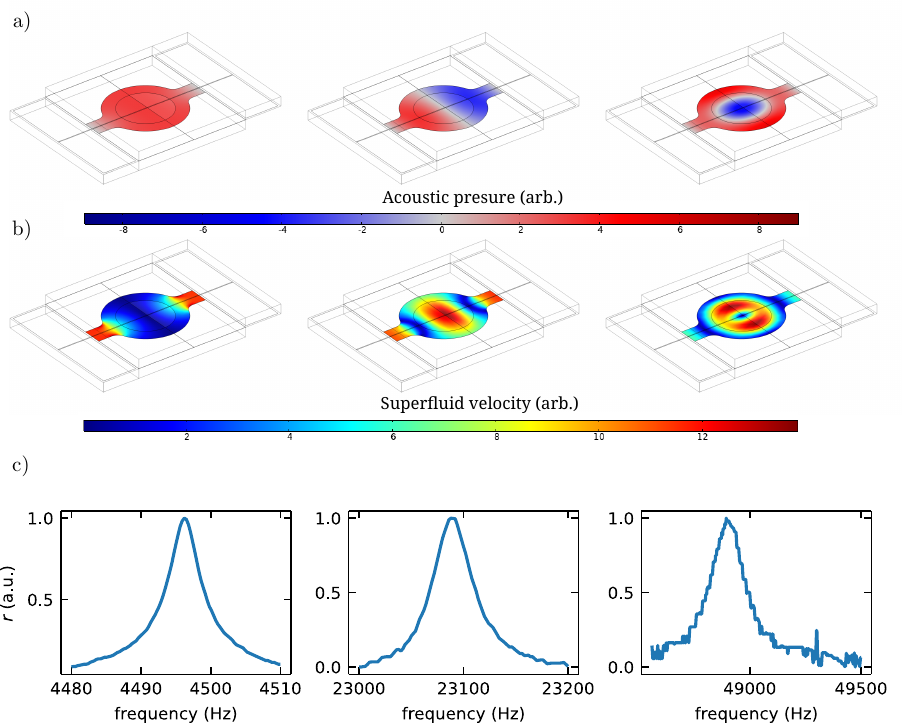}
    \caption{Acoustic modes inside the Helmholtz resonator. a) Finite element simulation of acoustic pressure profile for the first three visible modes. The bending of SiO$_2$ substrate is simulated alongside pressure acoustics in a coupled 3D geometry (the COMSOL "multiphysics" coupling). b) Superfluid velocity calculated from the same simulation. c) Experimentally observed resonances using EMIT, with detuning $\Delta = 0$ in order to observe the intrinsic mechanical linewidth. The signal from lock-in amplifier is offset to zero and normalized to 1 on resonance. See Tab.~\ref{tab:simulations} for calculated properties of the acoustic modes.}
    \label{fig:fig3}
\end{figure}

\begin{figure}
    \includegraphics[width=0.90\linewidth]{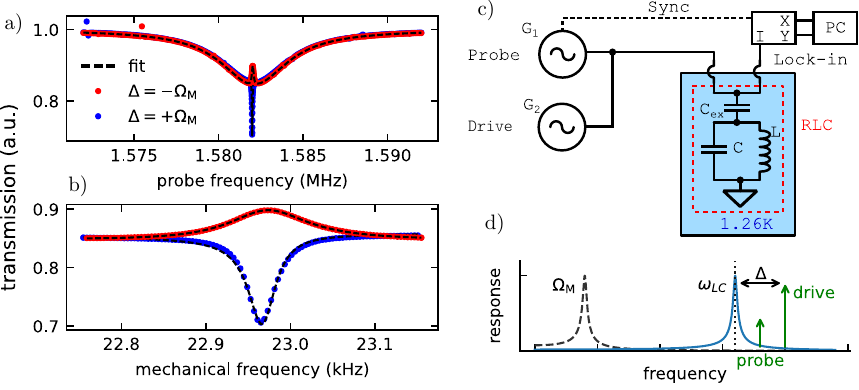}
    \caption{Electromechanically induced transparency and absorption. a) EMIT (red) and EMIA (blue) measurement of the coupled system. b) Zoom to acoustic peak on LC resonance. The $x$-axis is $|\omega_d - \omega_p|/2\pi$. c) Circuit diagram for measuring EMIT/A, the strong drive signal from G2 is added to weaker probe signal from G1 (which is measured using a lock-in amplifier). d) Relation of drive and probe frequencies to $\wlc$ and $\wm$.}
    \label{fig:fig4}
\end{figure}

Finite element modelling (FEM) of the superfluid motion in the nanofluidic device for several acoustic modes is shown in Fig.~\ref{fig:fig3}. To drive and detect the acoustic modes we use Electro-Mechanically Induced Transparency or Absorption (EMIT/A, circuit diagram in Fig.~\ref{fig:fig4}c), similar to OMIT/A \cite{weis_optomechanically_2010}, where a strong drive tone is set to a constant detuning $\Delta$ from the LC resonance $\omega_{d} =  \omega_{LC} + \Delta$, while a weaker probe tone sweeps a frequency range around $\omega_p = \omega_d \pm \wm$ (sign such that $\omega_p$ is close to $\omega_{LC}$), where $\wm$ is the frequency of the acoustic mode. As force on capacitor plates is proportional to the square of voltage $U_C^2$, a force is exerted on the difference frequency $\omega_p - \omega_d \sim \Omega_\mathrm{M}$. The mechanical motion coherently mixes with the drive tone at $\omega_d$, resulting in sidebands that interfere with $\omega_p$ observed using a lock-in amplifier, leading to mechanical peak appearing superimposed on the LC response, as shown in Fig.~\ref{fig:fig4}.

The first three identified acoustic modes, matched with the mode shape obtained from simulations, can be seen in Fig.~\ref{fig:fig3}. Not all of the acoustic modes are detectable, as the averaged deflection of electrodes caused by the pressure of the acoustic mode must be non-zero (compare with the electrode geometry shown in Fig.~\ref{fig:fig1}). The calculated parameters of the acoustic modes are shown in Tab.~\ref{tab:simulations}.

\begin{table}
    \centering
    \begin{tabular}{lcccccc}
        mode & $f_0$ (kHz) & $m_\mathrm{eff}$ (ng) & $y_\mathrm{ZPF}$ (fm) & $g_0$ (\textmu Hz) & measured $f_0$ (kHz) & $\partial y_
        \mathrm{el}/\partial y$ (-) \\\hline
         mode 1 & 5.003 & 267 & 2.5 & 0.283 & 4.492 &
         $\mathrm{7.66\times10^{-5}}$\\
         mode 2 & 24.47 & 588 & 0.76 & 0.237 & 23.09 & 
         $\mathrm{2.11\times10^{-4}}$\\
         mode 3 & 49.98 & 500 & 0.58 & 0.261 & 48.90 &
         $\mathrm{3.06\times10^{-4}}$ \\
    \end{tabular}
    \caption{Simulated properties of the acoustic modes shown in Fig.~\ref{fig:fig3}: $f_0$ -- resonance frequency, $m_\mathrm{eff}$ -- effective mass, $y_\mathrm{ZPF}$ -- zero-point fluctuations, $g_0$ -- vacuum coupling rate, measured $f_0$ -- experimentally measured mode frequency; $\partial y_
        \mathrm{el}/\partial y$ -- relative displacement of the device capacitor electrodes with respect to the in-plane maximum helium displacement}
    \label{tab:simulations}
\end{table}

\subsection{Electromechanical vacuum coupling rate}
The vacuum coupling \cite{Aspelmeyer2014} $g_0$ is given by
\begin{equation}
    g_0 = G y_\mathrm{ZPF},
\end{equation}
where $G = \dd\omega_{LC} / \dd y$ is the change of LC resonance frequency with equivalent harmonic oscillator displacement $y$, and
\begin{equation}
    y_\mathrm{ZPF} = \sqrt{\frac{\hbar}{2m_\mathrm{eff}\Omega_\mathrm{M}}}
\end{equation}
is the amplitude of zero-point fluctuations, where $m_\mathrm{eff}$ is the effective mass of the acoustic mode. Note that there is some freedom of choice regarding the $y$ normalization, which in turn gives the normalization of $m_\mathrm{eff}$. We follow the convention of \cite{Hauer2013}: The instantaneous local displacement of the acoustic mode is $\mathbf{y}(\mathbf{x},t)$, which can be decomposed into equivalent harmonic oscillator displacement $y(t)$ and the dimensionless mode shape $\mathbf{r}(\mathbf{x})$ 
\begin{equation}
	\mathbf{y}(\mathbf{x}, t) = \mathbf{r}(\mathbf{x}) y(t),
\end{equation}
where  $|\mathbf{r}(\mathbf{x})| \leq 1$, so $y(t)$ is taken as the acoustic displacement of the fluid element that moves the most. We follow \cite{Hauer2013} to obtain $m_\mathrm{eff}$ from numerical simulation, where superfluid helium is an acoustic medium coupled to flexible SiO$_{2}$, by integral over the spatial coordinate $\mathbf{x}$
\begin{equation}
    m_\mathrm{eff} = \int_V \rho (\mathbf{x}) |\mathbf{r} (\mathbf{x})|^2 \dd V,
\end{equation}
where integration is over both helium and SiO$_2$, hence the $\mathbf{x}$ dependence of local density $\rho$ and $\mathbf{r}(\mathbf{x})$ is the mode shape in Fig.~\ref{fig:fig3}. The calculated effective masses are listed in Tab.~\ref{tab:simulations}. The SiO$_2$ displacement is, however, three orders of magnitude smaller compared to helium acoustic displacement, and does not appreciably change the mode effective mass. 

The frequency pull parameter $G$ is calculated from the LC resonance frequency $\omega_{LC} = [L(C + C_\mathrm{ex})]^{-1/2}$ (see Appendix~\ref{app:circuit} for derivation of $\omega_{LC}$) as
\begin{equation}
G = \frac{\partial \omega_{LC}}{\partial y} = \frac{\partial \omega_{LC}}{\partial y_\mathrm{el}} \frac{\partial y_\mathrm{el}}{\partial y}.
\end{equation}
The average displacement of the electrode $y_\mathrm{el}$ is given by
\begin{equation}
\label{eq:yel}
y_\mathrm{el} = \frac{1}{S}\int_S y r_z(\mathbf{x}) \dd S = a y,
\end{equation}
where the integration runs over the electrode surface $S$ and we assume that $z$ direction is perpendicular to electrode surface. The normalized averaged electrode deflection $a = (\int_S r_z(\mathbf{x})\dd S)/S$ is obtained numerically (shown in Tab.~\ref{tab:simulations}), and we see that $\partial y_{el}/ \partial y = a$. For a parallel plate capacitor, $C = \varepsilon S/(D-2y_{el})$, we arrive at
\begin{equation}
G = \frac{C}{C_\mathrm{ex} + C}\frac{\omega_{LC}}{D} \frac{\partial y_{el}}{\partial y},
\end{equation}
where $C_\mathrm{ex} = 10$~pF and $C = 25$~pF are the external coupling and device capacitances, respectively (see Fig.~\ref{fig:fig2}), and $D=780$~nm is the electrode spacing. 

The numerically estimated effective masses, zero point fluctuations and vacuum coupling rates are shown in Tab.~\ref{tab:simulations}. Specifically, for mode 2 used in the rest of the paper we find
\begin{equation}
    \label{eq:g0-theory}
    g_{0,\mathrm{th}} = G y_\mathrm{ZPF} = 2\pi\times0.24 \; \mathrm{\mu Hz}.
\end{equation}

We extract the vacuum coupling rate from EMIT/A measurement of the mode 2, where the LC peak was scanned with a weak probe tone and a stronger drive tone was set at $\omega_d = \omega_{LC} \pm \Omega_\mathrm{M}$, shown in Fig. \ref{fig:fig4}. The vacuum coupling was obtained by fit of the transmission to \cite{Aspelmeyer2014}
\begin{equation}
    \label{eq:EMIT}
    t =  1 - \frac{\kappa_\mathrm{ex}}{2} \frac{\chi_{LC}(\omega)}{1 \mp g^2 \chi_\mathrm{mech}(\omega) \chi_{LC}(\omega)},
\end{equation}
where
\begin{equation}
    \label{eq:EMIT-chis}
    \chi_{LC}(\omega) = \frac{1}{\kappa/2 + i(\omega - \omega_{LC})} \; , \; \chi_\mathrm{mech}(\omega) = \frac{i}{\Omega_\mathrm{M} \pm (\omega - \omega_d) + i\Gamma_\mathrm{M}/2},
\end{equation}
and $\omega$ is the probe frequency. Fit to \eqref{eq:EMIT} gives the enhanced coupling rate $g = \sqrt{n_{LC}} g_0$ due to strong \emph{drive} tone, from which $g_0$ can be extracted if the number of circulating photons (given by drive tone) is known. This is estimated as
\begin{equation}
    \label{eq:nlc}
     n_{LC} = \frac{E}{\hbar \omega_{LC}} = \frac{\kappa_\mathrm{ex}}{\kappa^2 /4 + \Delta^2} \frac{R_D}{8 (R_S + R_D) R_S \hbar \omega_{LC}} U^2,
\end{equation}
where $\Delta = \omega_d - \omega_{LC}$ is the drive detuning, $R_D$ is the lock-in input impedance, $R_S$ is the generator output impedance and $U$ is the open-load voltage set on the drive generator, see full derivation in the Appendix~\ref{app:circuit}. The vacuum coupling rate extracted from the fit to experimental (detuning on red sideband only) data is
\begin{equation}
    g_0 = 2\pi\times0.19(7) \; \mathrm{\mu Hz},
\end{equation}
which is in reasonably good agreement with theoretically calculated coupling \eqref{eq:g0-theory}. 

\section{Dynamical backaction}
\label{sec:backaction}
Damping $\Gamma_\mathrm{M}$ as well as resonance frequency $\Omega_\mathrm{M}$ of the acoustic mode will be modified by dynamical backaction \cite{Aspelmeyer2014} depending on drive tone strength and detuning $\Delta$ from the LC resonance, i.e. optomechanical damping ($\Delta < 0$) or anti-damping ($\Delta > 0$)
\begin{equation}
    \label{eq:gamma_theory}
    \delta\Gamma_{M} = g^2 \left[ \frac{\kappa}{(\Delta + \Omega_\mathrm{M})^2 + \kappa^2/4} - \frac{\kappa}{(\Delta - \Omega_\mathrm{M})^2 + \kappa^2/4} \right],
\end{equation}
and optical spring effect
\begin{equation}
    \label{eq:domega_theory}
    \delta\Omega_\mathrm{M} = g^2 \left[ \frac{\Delta + \Omega_\mathrm{M}}{(\Delta + \Omega_\mathrm{M})^2 + \kappa^2/4} + \frac{\Delta - \Omega_\mathrm{M}}{(\Delta - \Omega_\mathrm{M})^2 + \kappa^2/4} \right].
\end{equation}

\begin{figure}
    \centering
    \includegraphics{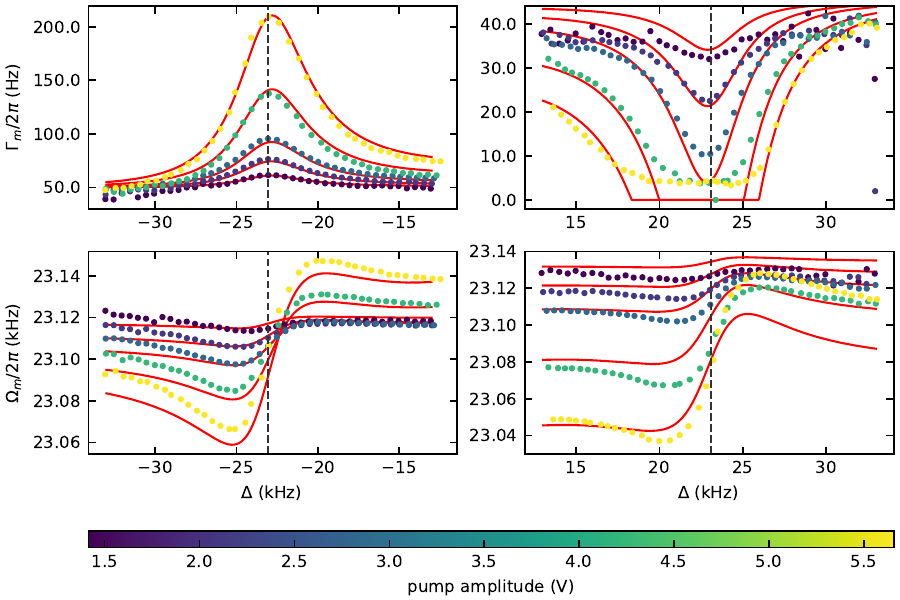}
    \caption{Mechanical peak parameters as a a function of drive detuning. Red lines are fits to \eqref{eq:gamma_theory} (first row) and \eqref{eq:domega_theory} (second row). Both electromechanical damping and amplification and the electromechanical spring effect are apparent. The dashed vertical lines show $\pm\wm$}
    \label{fig:backaction-all}
\end{figure}

We observe the dynamical backaction using standard EMIT/A (i.e., circuit diagram in Fig.~\ref{fig:fig4}) at increased drive amplitude at 1.26~K. The mechanical frequency $\wm$ and damping $\Gamma_\mathrm{M}$ are determined by fitting mechanical peak alone (without LC background) to mechanical resonance in \eqref{eq:EMIT-chis}, and are shown in Fig.~\ref{fig:backaction-all} for red and blue detuning and different amplitudes of the EMIT/A drive. We find good agreement with \eqref{eq:domega_theory}, \eqref{eq:gamma_theory} (red lines in Fig.~\ref{fig:backaction-all}). Common decay rates $\kappa$ and $\kappa_\mathrm{ex}$ were used for all fits at a given drive voltage $U_d$. We attribute the slight imperfections of the fits (more visible on the blue sideband) to nonlinearity of the LC circuit as seen in Fig.~\ref{fig:fig2}. At $\Delta = +\wm$ at sufficiently high drives, the mechanical linewidth becomes zero according to the linearized theory \cite{Aspelmeyer2014} (seen in Fig.~\ref{fig:backaction-all}), at which point the system begins to self-oscillate.

\begin{figure}
    \centering
    \includegraphics[width=0.9\linewidth]{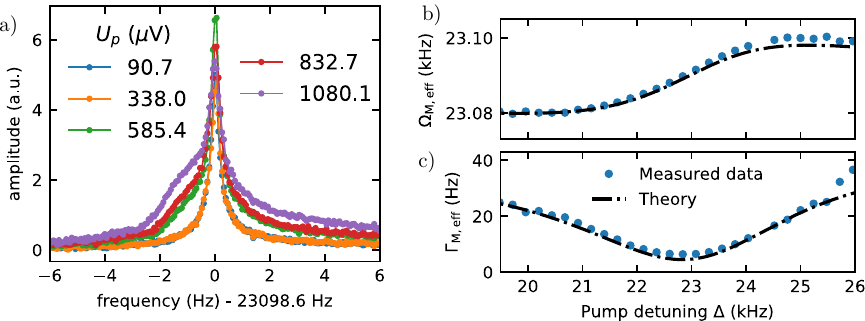}
    \caption{Pulsed EMIA. a) Probe power dependence of the mechanical peak in the pulsed EMIA spectrum at fixed drive $U_d \approx 2.06$~V. As the probe power (i.e., AM sideband amplitude) increases, the superfluid acoustic mode becomes non-linear, presumably due to turbulence. b) Dynamical backaction -- optical spring effect (upper) and optomechanical anti-damping (lower) as a function of drive detuning for fixed $U_d \approx 0.88$~V and $U_p \approx 560$~\textmu V. The theory lines are given by \eqref{eq:domega_theory},\eqref{eq:gamma_theory} with $g$ given by \eqref{eq:g0-theory} and \eqref{eq:nlc} (i.e., no fitting).}
    \label{fig:threetone}
\end{figure}

Restricting the attention to the blue-detuned drive $\Delta > 0$, we implemented a pulsed EMIA measurement with results shown in Fig.~\ref{fig:threetone}. The drive is tuned to $\omega_d \approx \wlc + \wm$, from which the probe sidebands are derived by amplitude modulation (AM) at $\wm$ resulting in AM sidebands at $\wlc$ and $\wlc + 2\wm$. The scheme is similar to three-tone measurements \cite{Kashkanova2016,Kashkanova2017,spence_three-tone_2023}, however, since the system is sideband resolved, the higher AM sideband at $\Delta \approx 2\wm$ is effectively filtered out by the LC circuit and the measurement reduces to ordinary EMIA.

Based on \cite{doolin_integrated_2019,shook_stabilized_2020}, for the the modulating signal we use a chirped pulse of the form $s(t) \propto \sin\left[2\pi (f_i t + \frac{1}{2}(f_f - f_i)t^2/\tau)\right]$, where $f_{i,f}$ are the initial and final frequencies of the pulse and $\tau$ is the duration of the pulse. The pulse limit frequencies $f_{i,f}$ span $\wm/2\pi$ and $\tau \approx 1$~s. The transmitted signal is down-mixed with $\omega_d$, filtered and then down-mixed again using a Stanford Research SR830 lock-in amplifier with (arbitrary) reference set close to $\wm$ and with bandwidth (i.e., inverse lockin time constant) set to exceed the mechanical linewidth. The output quadratures $x$ and $y$ were recorded using fast data acquisition, triggered to the beginning of the chirped pulse, as a complex signal $z(t) = x(t) + i y(t)$. The final EMIA spectrum is calculated as $\chi_\mathrm{M}(\omega) = \mathcal{F}[z] / \mathcal{F}[s]$, with $\mathcal{F}$ denoting the Fourier transform.

The narrow-band amplification of the lockin helps with suppressing spurious noise peaks before the final digitisation of the signal, which allows for relatively rapid measurement of the the mechanical spectra. The pulsed nature of the measurement also avoids exciting the acoustic motion to high velocities where the mechanical resonance becomes nonlinear, likely due to turbulence \cite{novotny_critical_2025,novotny_decay_2026}. Mechanical nonlinearity, shown in Fig.~\ref{fig:threetone}a is observed only at high pulse amplitudes (i.e., the AM sidebands amplitude), which gives us approximately one decade in pulse amplitude where the signal to noise is sufficient and no nonlinearity is present. The spring effect and optomechanical heating (Fig.~\ref{fig:threetone}b,c) behave similarly to the non-pulsed measurement and in accordance with \eqref{eq:domega_theory},\eqref{eq:gamma_theory}, as expected since the dynamical backaction depends primarily on the strong drive power, which is not pulsed but rather applied continuously as in the non-pulsed measurements.

\section{Rotation-induced dissipation}
\label{sec:vortices}

The acoustic mode is damped by vortices present in the device, created either by external rotation \cite{novotny_detection_2024,dwivedi_dynamics_2024} or by turbulence \cite{novotny_critical_2025}. We can write acoustic damping in the form
\begin{equation}
    \Gamma_{\mathrm{M},\mathrm{tot}} = \Gamma_\mathrm{M} + \delta\Gamma_\mathrm{M} + \Gamma_L,
\end{equation}
where $\Gamma_\mathrm{M}$ is the intrinsic linewidth given mainly due to normal fluid viscosity \cite{Souris2017}, $\delta\Gamma_\mathrm{M} < 0$ is due to the backaction, and $\Gamma_L = \hat\alpha \kappa L$ is the dissipation due to quantized vortices, where $L$ is the area density of vortex lines perpendicular to the flow velocity \cite{novotny_detection_2024}, $\kappa = h/m_4 \approx 9.97 \times 10^{-7}$~m$^2$/s ($m_4$ is the helium atom mass) is the quantum of circulation and $\hat\alpha$ is the mutual friction constant modified by vortex-surface interaction in confined geometry \cite{novotny_decay_2026}.

The equilibrium density of vortices in a container rotating with angular velocity $\theta$ parallel to the axis of rotation is $L_\mathrm{eq} = 2\theta/\kappa$ \cite{peretti_direct_2023}. For the bare mutual friction constant $\alpha$ (i.e., not enhanced by the interaction with walls \cite{novotny_decay_2026}) and a typical experimentally achievable value of $\theta= 1.6$~rad~s$^{-1}$, the additional damping is $\Gamma_\mathrm{eq} = 2\theta\alpha = 2\pi\times0.017$~Hz. Resolving this small change requires significant amount of averaging \cite{novotny_detection_2024}, which essentially precludes observing dynamical changes in the vortex system. However, the dynamical backaction allows to narrow the acoustic mode from $\Gamma_\mathrm{M} = 2\pi\times 55$~Hz to $\Gamma_{\mathrm{M},\mathrm{eff}} = \Gamma_\mathrm{M} + \delta\Gamma_\mathrm{M} \approx 2\pi\times0.5$~Hz for the brief duration of the EMIA pulse. This amplification of the acoustic motion also leads to enhancement of effective $\hat\alpha$, which is strongly velocity-dependent \cite{novotny_decay_2026}.

Therefore, to detect quantized vortices, we employed the pulsed EMIA scheme described above. First, $\Delta$ was tuned to the lowest point $\delta\Gamma_\mathrm{M}$ curve shown in Fig.~\ref{fig:threetone}c, and then $U_d$ was increased until the total linewidth $\Gamma_{\mathrm{M},\mathrm{eff}} < 2\pi\times0.8$~Hz. At fixed $\Delta$, $U_d$ and EMIA pulse parameters, the relationship $\Gamma_{\mathrm{M},\mathrm{eff}} (\wm)$ was calibrated by allowing the temperature to drift around the experimental setpoint $T\approx 1.26$~K when the device was in a vortex-free state, see Appendix~\ref{app:damping} (temperature in cryostat could not be PID regulated while rotating, and a $\Delta T< 1~\mathrm{mK}$ drift was observed). Note that by "vortex-free", we mean a high-$Q$ reproducible state of the dissipation; it is possible that this state still contains trapped vortices which cannot be removed by the techniques described below and any additional damping is relative to this reproducible background. The vortex-induced damping was then calculated as $\Gamma_L = \Gamma_{\mathrm{M},\mathrm{tot}} - \Gamma_{\mathrm{M},\mathrm{eff}}(\wm)$, where $\Gamma_{\mathrm{M},\mathrm{tot}}$ and $\wm$ were directly obtained from a fit to the EMIA peak.

\begin{figure}
    \centering
    \includegraphics[width=0.9\linewidth]{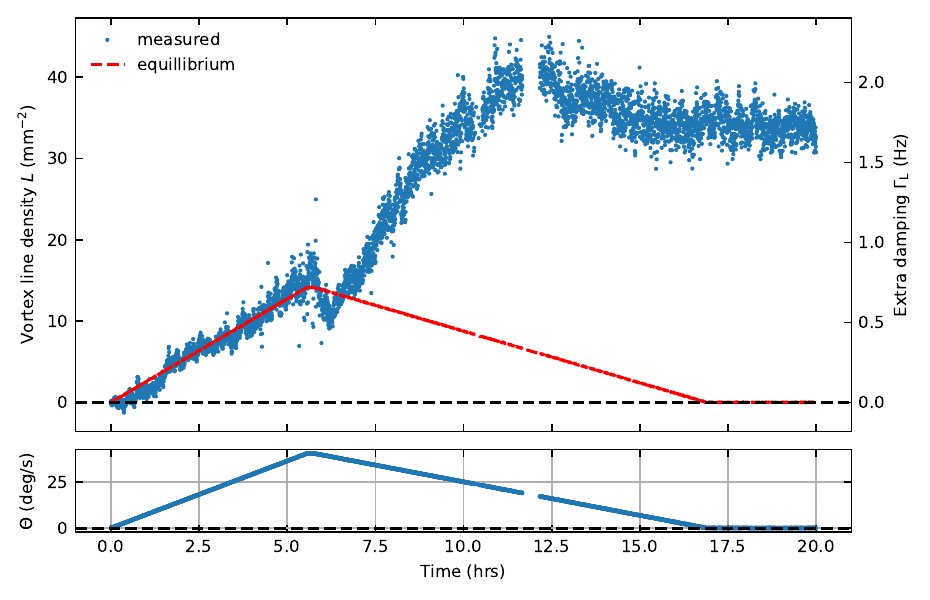}
    \caption{(top) Rotation-induced dissipation. a) Additional dissipation $\Gamma_L$, with respect to the calibrated background (see Appendix~\ref{app:damping}) as a function of time during ramped angular velocity. (bottom) Angular velocity as a function of time. The red line in the top panel is a fit of the initial linear part of the dissipation to equilibrium vortex line density with adjustable effective mutual friction.}
    \label{fig:feynman-accel}
\end{figure}

The response of $\Gamma_L$ to constant slow acceleration of the rotating cryostat is shown in Fig.~\ref{fig:feynman-accel}. In the initial acceleration, the added damping increases approximately linearly in time. After reaching 40~deg/s, acceleration is stopped, and the angular velocity is ramped down linearly over a period of 12 hours. During this breaking period we see, surprisingly, an increase in damping and the device finally settles to a state of increased dissipation after the cryostat comes to a complete stop.

The red line in Fig.~\ref{fig:feynman-accel} is a fit of $\Gamma_L$ to the initial linear increase with adjustable $\hat\alpha$. We find that $\hat\alpha \approx 15\alpha$, where $\alpha$ is the bare mutual friction constant \cite{donnelly_observed_1998}. This is slightly higher that the maximum of velocity-dependent $\hat\alpha$ in ref.~\cite{novotny_decay_2026}, however, there low-frequency flows were considered. In the present case, the flow oscillation frequency $\wm$ exceeds the bulk decay rate of Kelvin waves $\gamma_\mathrm{KW} \approx \alpha\omega_\mathrm{KW} \approx 2\pi \times 5$~kHz, with $|\omega_\mathrm{KW}| \approx \pi\kappa/4D^2 \left[\log\left(2D / \pi a \right) - 0.5772\right] \approx 2\pi\times147$~kHz \cite{barenghi_thermal_1985}, $a\approx 1.5$~\AA~ the vortex core and $\alpha\approx 0.034$ the bulk mutual friction at 1.3~K \cite{donnelly_observed_1998}. This will likely lead to the breakdown of the 2D approximation used in \cite{novotny_decay_2026} and more frequent re-pinning with the surface \cite{Donev2001a} and therefore to higher dissipation. The extension of the effective mutual friction to higher frequencies will be the subject of future work.

\begin{figure}
    \centering
    \includegraphics{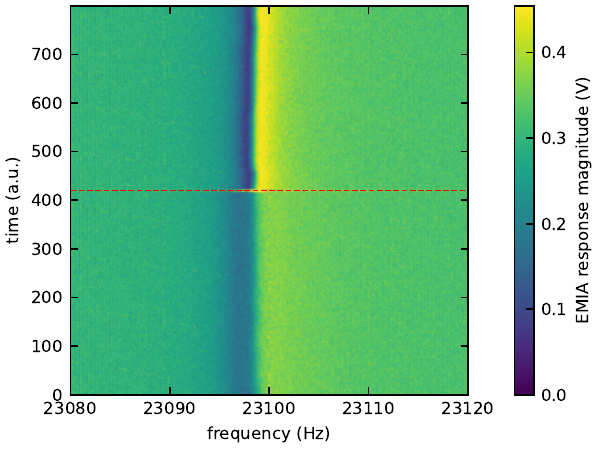}
    \caption{The effect of annealing on the spectrum measured by the pulsed EMIA. Each horizontal line represents one measurement of the mechanical spectrum. The annealing pulse was applied at the time indicated by the dashed horizontal line.}
    \label{fig:annealing}
\end{figure}

The increased dissipation state after stopping the cryostat is stable in time -- no change was observed over the period of 48 hours when the system was monitored with EMIA pulses, but left undisturbed otherwise. It can, however, be reverted to the original high-$Q$ state in several ways: briefly driving the system over the self-oscillation instability using high $U_d$ (drive amplitude); using a strong pulse $U_p$; or tapping the cryostat with a hammer. Typically multiple taps are required, as the damping is reduced only partially (but it is always reduced). We refer to the non-hammer-based methods as \emph{annealing}. The effect of the annealing is shown in Fig.~\ref{fig:annealing}, which shows the evolution of the EMIA spectrum in time. Immediately after annealing, the $Q$ increases approximately to its vortex-free level obtained during calibration (see Appendix~\ref{app:damping}). Further annealing pulses have no effect. 

\begin{figure}
    \centering
    \includegraphics[width=0.9\linewidth]{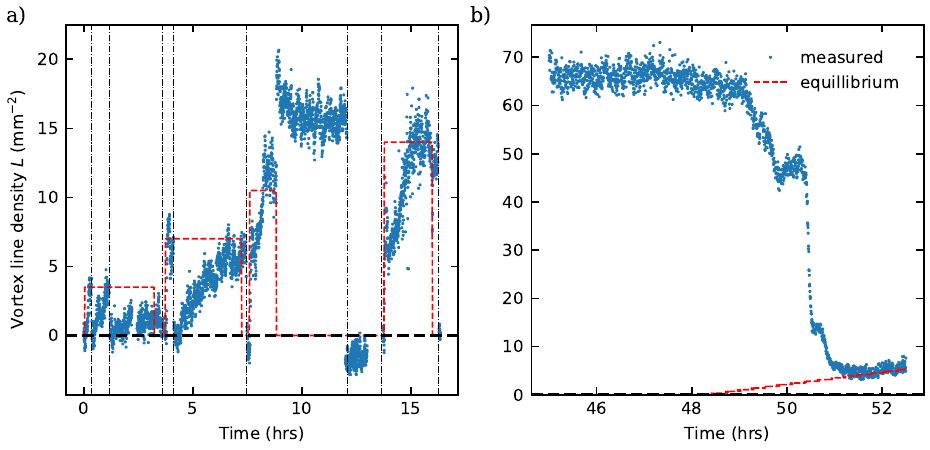}
    \caption{Effects of rotation and annealing. a) As Fig.~\ref{fig:feynman-accel}, the red dashed line represents expected damping based on equilibrium vortex line density given by the rotation and $\hat\alpha$ given by the linear fit in Fig.~\ref{fig:feynman-accel}. The vertical black dash-dotted lines indicate annealing pulses. The equilibration at a given angular velocity is history dependent and can take several hours. The annealing pulses, however, consistently remove vortices from the system. b) Vortex avalanche. Reversing the rotation direction with respect to the state in which the vortices were trapped results in sudden decrease of the vortex density to levels expected for the equilibrium density. Before the collapse, the vortex line density was stable for about 12 hours.}
    \label{fig:feynman-stepped}
\end{figure}

The general behaviour of the dissipation is complex, history-dependent and to some extent stochastic, of which we show two examples in Fig.~\ref{fig:feynman-stepped}. In Fig.~\ref{fig:feynman-stepped}a we show an example 16-hour continuous measurement sequence with several rotation velocities and annealing pulses. The dissipation tends to respond quickly to sudden changes in rotation velocity (but can take minutes to hours to settle) and it tends to increase when rotation velocity is decreased. Annealing pulses reduce dissipation even during rotation. The dissipation takes several hours to recover in steady rotation after an annealing pulse, however. In Fig.~\ref{fig:feynman-stepped}b we show an avalanche-like reduction in damping. In this case, vortices trapped by rotation and slow deceleration are used as the initial condition. At time $t\approx 48$~h we start rotation of the cryostat in opposite direction to the previous steady rotation. This leads to sudden collapse of the vortex density, which follows the new equilibrium density afterwards.

The observed behaviour is generally consistent with vortices pinned on the surface irregularities. Indeed, similar behaviour was recently observed in $^3$He confined in rotating aerogel \cite{autti_exceeding_2020}, where total vorticity in the system is reduced by nucleating additional oppositely oriented vortices upon deceleration of the system (although it should be kept in mind that, unlike in $^3$He, the surface irregularities in our case exceed the vortex core size by at least an order of magnitude). The annealing pulse in our case likely exceeds the critical velocity for depinning \cite{schwarz_three-dimensional_1985}, which ejects the vortices from areas of highest velocity, where the dissipation of the acoustic mode is the most sensitive to vortices.

Finally, we note that we have repeated a subset of the experiments without the LC resonator and both $\hat\alpha \approx 15\alpha$, which depends on flow velocity amplitude (but is roughly the same for comparable peak velocities between the configurations), and the annealing effect was reproduced. Therefore, these effects are not related either to the parametric readout or backaction.

\section{Conclusions}

We have demonstrated a nanofluidic acoustic resonator parametrically coupled to a superconducting LC tank circuit operating in the sideband-resolved regime. We observed the mechanical motion using electromechanically-induced transparency or absorption (EMIT/A), which was also used to observe dynamical backaction -- optical spring effect and optomechanical damping and amplification. Using pulsed EMIA, we found complex and history-dependent response of the acoustic dissipation to rotation of the system, which we attribute to vortex trapping on the surface roughness, although the understanding is at present at a qualitative rather than quantitative level.

The sensitivity to vorticity and rotation coupled with strong pinning make the system highly promising for the study of vortex avalanches important in the physics of pulsar glitches. Indeed, laboratory superfluid analogies to neutron star spindown were considered in the past \cite{tsakadze_properties_1980}. The present experiment closely resembles recent simulations using the Gross-Pitaevskii equation \cite{liu_vortex_2025} and point vortex model \cite{autti_superfluid_2026}. Although in our case the pinning appears to be too strong for glitches in vortex number to appear on simple deceleration, an avalanche-like rapid reduction in vortex damping was seen by either sufficiently exciting the acoustic mode (annealing) or negatively-oriented rotation of the cryostat.

Additionally, and somewhat speculatively, the fast oscillation of the acoustic mode appears to mobilize the vortices without necessarily inducing turbulence, which, on long timescales, provides in-situ dynamic control over vortex pinning strength. The velocity profile of the acoustic mode then translates to large-scale inhomogeneities in pinning strength. Such ``vortex traps'' are believed to be important for reproducing the observed statistics of pulsar glitches \cite{viswanathan_pulsar_2026,alpar_postglitch_1996}. This potential analogy with pulsar glitches will be explored in future studies.

\section*{Acknowledgements}

We are grateful to L. Skrbek, M. A. Alpar and M. Urbanec for fruitful discussions facilitated by COST Action CA24139, Superfluid Condensates in Astrophysics and Laboratory Experiments (SCALES), supported by COST (European Cooperation in Science and Technology).

The work was supported by the Czech Science Foundation under GA\v{C}R 25-16386S. M.T. acknowledges support by the Charles University under GAUK 434126. CzechNanoLab projects LM2023051 and LNSM-LNSpin funded by MEYS CR are also gratefully acknowledged for the financial support of the sample fabrication at CEITEC Nano Research Infrastructure and LNSM at FZU AV\v{C}R. AFM characterization was performed in MGML laboratories supported by the program of Czech Research Infrastructures (project no. LM2023065).

\section*{Data}
Data and code for all figures are available in Zenodo repository \href{https://doi.org/10.5281/zenodo.22144520}{10.5281/zenodo.22144521}.


\section*{Appendix}
\appendix

\section{Lumped element LC analysis}
\label{app:circuit}
At frequencies close to circuit resonance of about 1.6 MHz, its electromagnetic properties are well described in terms of lumped circuit elements. The scheme of the circuit for this analysis is in Fig.~\ref{fig:fig2}b). Our goal is to obtain voltage on detector $U_D$, external coupling $\kappa_\mathrm{ex}$, intrinsic line width $\kappa_0$ total line width $\kappa = \kappa_0 + \kappa_\mathrm{ex}$, resonance frequency $\omega_{LC}$ and photon number $n_{LC}$ from known parameters: device capacitance $C$, coupling capacitance $C_\mathrm{ex}$, coil inductance $L$, coil effective series resistance $R_L$, source impedance $R_S$ and detector impedance $R_D$. 

First we calculate the impedance $Z_{LC}$ of the coupled circuit at angular frequency $\omega$
\begin{equation}
    \label{ZLC}
    Z_{LC} = \frac{1}{i\omega C_\mathrm{ex}} + \left( \frac{1}{R_L + i\omega L}  + i\omega C\right)^{-1} = \frac{1 - \omega^2 L (C+C_\mathrm{ex}) + i\omega R_L (C + C_\mathrm{ex})}{i\omega C_\mathrm{ex} (1  - \omega^2 L C + i\omega R_L C )}. 
\end{equation}
Maximal current will flow to circuit as $Z_{LC} \rightarrow 0$, so resonance will occur at minimal $Z_{LC}$. Minimizing real part of numerator gives us the LC resonance frequency
\begin{equation}
    \omega_{LC} = \frac{1}{\sqrt{L(C+C_\mathrm{ex})}},
\end{equation}
and substituting for $\omega_{LC}$ and reorganizing we get
\begin{equation}
    Z_{LC} = \frac{\omega_{LC}^2 - \omega^2 + \frac{i\omega R_L}{L}}{i\omega C_\mathrm{ex} \left(\omega_{LC}^2 - \omega^2 \frac{C}{C + C_\mathrm{ex}} +i \omega \frac{R_L}{L}\frac{C}{C+C_\mathrm{ex}} \right)}.
\end{equation}
Introducing the capacitance ratio
\begin{equation}
    r = \frac{C_\mathrm{ex}}{C},
\end{equation}
and what we will later show to be intrinsic linewidth
\begin{equation}
    \kappa_0 = \frac{R_L}{L},
\end{equation}
we rewrite $Z_{LC}$ as
\begin{equation}
    Z_{LC} = \frac{\left(\omega_{LC}^2 - \omega^2 + i\omega \kappa_0\right)(1+r)}{i\omega C_\mathrm{ex}\left((1+r)\omega_{LC}^2 - \omega^2 + i\omega \kappa_0\right)}.
\end{equation}
Substituting 
\begin{equation}
    \omega = \omega_{LC} + \Delta,
\end{equation}
assuming $\Delta \ll \omega_{LC}$ as is the case in experiment, yields
\begin{equation}
    Z_{LC} \approx \frac{(- 2\Delta +i\kappa_0)(1+r)}{i\omega_{LC}C_\mathrm{ex}(r\omega_{LC} + i\kappa_0)}.
\end{equation}
The voltage on the detector is 
\begin{equation}\label{Ud_full}
    U_D = U \left(1 - \frac{R_S}{R_S + \left(Z_{LC}^{-1} + R_D^{-1}\right)^{-1}}\right),
\end{equation}
where $U$ is the voltage on the generator. This reduces to
\begin{equation}
    U_D = U \frac{R_D}{R_S + R_D} 
\end{equation}
off resonance ($|\Delta| \gg \kappa_0$), where $Z_{LC}^{-1} \sim 0$. The resonance curve will then have the form
\begin{equation}\label{Ud_u}
    U_D = U\frac{R_D}{R_S + R_D}\bigg(1 - u \bigg),
\end{equation}
where $0<u<1$. Comparing Eqs.~\eqref{Ud_full} and \eqref{Ud_u} we get
\begin{equation}
    u = \frac{R_S R_D}{R_S R_D + (R_S + R_D) Z_{LC}} = \frac{1}{2}\frac{i \kappa_\mathrm{ex}}{i(\kappa_0 + \kappa_\mathrm{ex})/2 - \Delta},
\end{equation}
where we introduced
\begin{equation}
   \kappa_\mathrm{ex} = \frac{ R_S R_D\omega_{LC} C_\mathrm{ex} \left(i \kappa_0 + r\omega_{LC} \right)}{(R_S + R_D)(1+r)} \sim \frac{ R_S R_D\omega_{LC}^2 C_\mathrm{ex} r}{(R_S + R_D)(1+r)}.
\end{equation}
The detector voltage can finally be written as ($\Delta = \omega - \omega_{LC}$)
\begin{equation}
    U_D = U \frac{R_D}{R_S + R_D}\left( 1 - \frac{1}{2}\frac{i \kappa_\mathrm{ex}}{i(\kappa_0 + \kappa_\mathrm{ex})/2 - \Delta}\right),
\end{equation}
which justifies the choices of $\kappa_0$ and $\kappa_\mathrm{ex}$. To obtain $n_{LC}$ we need the circulating energy in the LC circuit, which is a sum of peak energy on coupling capacitance and Helmholtz resonator capacitance
\begin{equation}
   E = E_C + E_\mathrm{ex} \;\; , \;\; E_\mathrm{ex} = \frac{1}{2}C_\mathrm{ex}|U_\mathrm{ex}|^2 \;\; , \;\; E_C = \frac{1}{2} C |U_C|^2.
\end{equation}
To obtain the energies, we need $U_\mathrm{ex}$ and $U_C$:
\begin{equation}
   U_\mathrm{ex} = \frac{U_D}{Z_{LC}}\frac{1}{i\omega_{LC} C_\mathrm{ex}} \;\; , \;\;  U_C = U_D\left(1 - \frac{1}{i\omega_{LC} C_\mathrm{ex} Z_{LC}}\right).
\end{equation}
Inserting $U_D$ in $U_\mathrm{ex}$ we get
\begin{equation}
   U_\mathrm{ex} =  U \frac{R_D}{R_S + R_D}\frac{r\omega_{LC}/2 + i\kappa_0/2}{(i(\kappa_0 + \kappa_\mathrm{ex})/2 - \Delta)(1+r)} = \frac{U}{i\omega_{LC} C_\mathrm{ex}R_S}\frac{i\kappa_\mathrm{ex}/2}{i(\kappa_0 + \kappa_\mathrm{ex})/2 - \Delta}.
\end{equation}
A similar procedure can be repeated for $U_C$ to arrive at
\begin{equation}
   U_{C} =  U \frac{R_D}{R_S + R_D}\frac{-(r\omega_{LC}/2 + ir\kappa_0/2)}{(i(\kappa_0 + \kappa_\mathrm{ex})/2 - \Delta)(1+r)} = \frac{U}{i\omega_{LC} C_\mathrm{ex}R_S}\frac{-i\kappa_\mathrm{ex}/2}{i(\kappa_0 + \kappa_\mathrm{ex})/2 - \Delta}.
\end{equation}
Inserting into the equation for $E$ we get, after simplifying 
\begin{equation}
    E = \frac{\kappa_\mathrm{ex}}{\kappa^2 /4 + \Delta^2} \frac{R_D}{8 (R_S + R_D) R_S} U^2
\end{equation}
for the energy on resonance. From there, photon number $n_{LC}$ is just
\begin{equation}
    n_{LC} = \frac{E}{\hbar \omega_{LC}} = \frac{\kappa_\mathrm{ex}}{\kappa^2 /4 + \Delta^2} \frac{R_D}{8 (R_S + R_D) R_S \hbar \omega_{LC}} U^2.
\end{equation}
When two signal generators are connected in parallel, new $R'_D$ reads as
\begin{equation}
    R'_D = \left(R_D^{-1} + R_S^{-1}\right)^{-1}
\end{equation}
where $R_D$ is detector impedance, and $R_S$ is the source generator internal impedance. 

\section{Damping calibration}
\label{app:damping}
Before rotating, we perform a calibration to compensate degraded temperature stability during rotation. This degraded stability is due to motor noise interfering with the resistance bridge used to measure resistive thermometers and due to centripetal force affecting the readout of saturated vapour pressure. Therefore, in a typical rotating measurement, the temperature was not actively stabilised by a PID controller and drifted slightly about the non-rotating setpoint of 1.26~K.

\begin{figure}
    \centering
    \includegraphics[width=\linewidth]{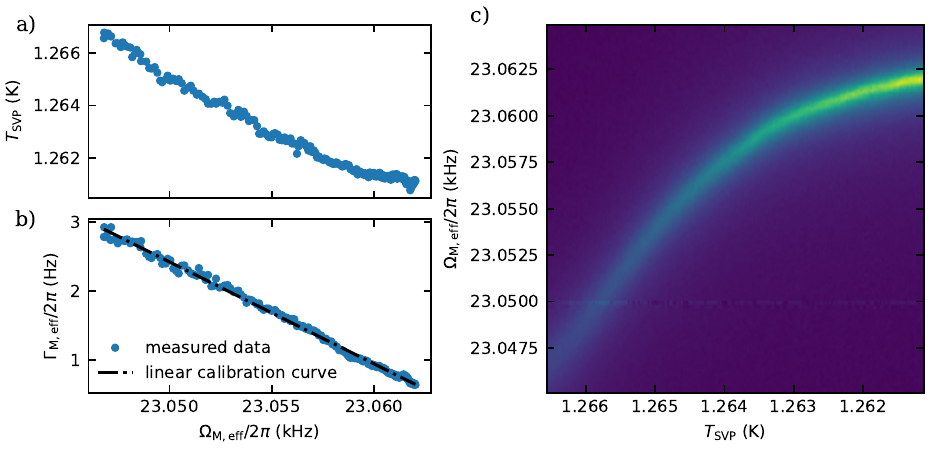}
    \caption{Calibration of the baseline damping. a,b) Temperature and acoustic linewidth $\Gamma_\mathrm{M}$ as a function of the resonance frequency $\wm$ at fixed EMIA drive. c) Change in the acoustic response with drifting temperature.}
    \label{fig:calibration}
\end{figure}

To compensate for this, we measure the acoustic resonance in a non-rotating, vortex-free (reproducibly annealed) state with fixed $\Delta$ and $U_d$ while allowing the temperature to drift in approximately 10~mK window about the setpoint. The mechanical spectra with decreasing temperature are shown in Fig.~\ref{fig:calibration}c, and the obtained curve of $\Gamma_{\mathrm{M},\mathrm{eff}}$ as a function of $\wm$ is shown in Fig.~\ref{fig:calibration}b. This was fit to a linear dependence and used to obtain $\Gamma_L = \Gamma_{\mathrm{M},\mathrm{tot}} - \Gamma_{\mathrm{M},\mathrm{eff}}(\wm)$, where $\Gamma_{\mathrm{M},\mathrm{tot}}$ and $\wm$ are obtained from a fit to an EMIT pulse which measures vortex line density.

\begin{figure}
    \centering
    \includegraphics[width=\linewidth]{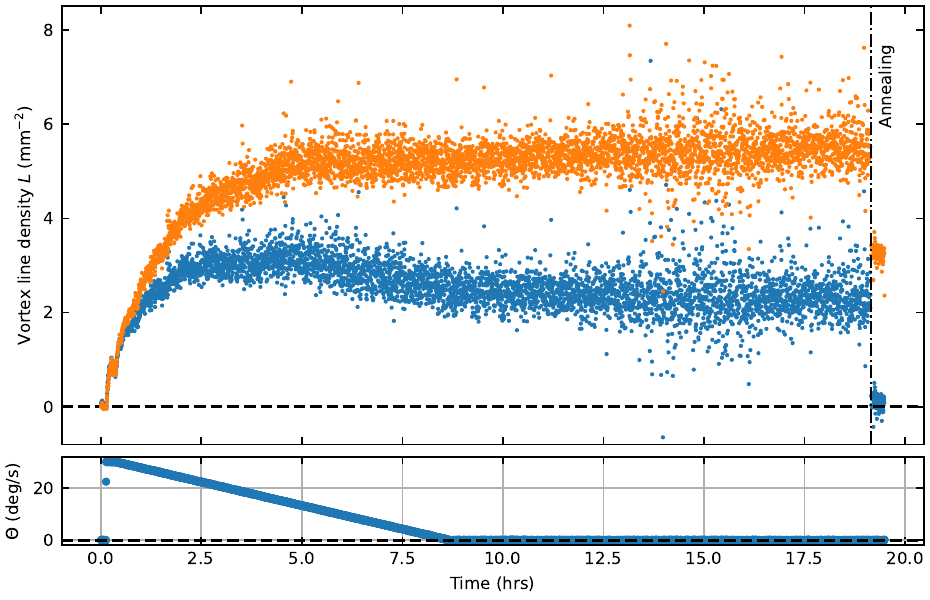}
    \caption{(top) Additional damping due to vortices created with rotation, and (bottom) rotation speed $\Theta$, as a function of time. The lower (blue) set of points is obtained by subtracting calibration curve from $\Gamma_\mathrm{M,eff}$ obtained from fit, the upper (orange) points by subtracting the unperturbed linewidth from the start of the sequence. At the end of the sequence, with stationary cryostat, quantized vortices were removed from the device by driving the mode into parametric instability. The calibrated data correctly show zero additional damping.}
    \label{fig:calibration-example}
\end{figure}

The effect of the calibration is illustrated in Fig.~\ref{fig:calibration-example}. The acoustic linewidth was measured as a response to rapid acceleration of the cryostat to $30$~deg/s, followed by a long deceleration to full stop over the span of approximately 8 hours. The two traces in Fig.~\ref{fig:calibration-example} show the excess damping $\Gamma_L$ either with respect to the initial state or with respect to the calibrated $\Gamma_\mathrm{M}(\wm)$. After approximately 20 hours, neither trace returns to the original state. However, using annealing (Sec.~\ref{sec:vortices}) by increasing the drive amplitude $U_d$ to achieve self oscillations, we see that without the calibration, almost 4 Hz of $\Gamma_L$ was due to drift, while calibrated $\Gamma_L$ reverts to zero, as expected.

\end{document}